\documentclass[reprint,aps,prc,nofootinbib,showkeys]{revtex4-1}

\usepackage{amsmath, amsthm, amsfonts, amssymb, xfrac, mathrsfs, txfonts}
\usepackage{graphicx,color,xcolor}
\usepackage{bm}	
\usepackage{vector} 
\usepackage{syntonly}
\usepackage{cancel,slashed}
\usepackage{hyperref}

\hypersetup{colorlinks,linkcolor=blue,citecolor=orange,urlcolor=teal}
\usepackage[capitalize]{cleveref}
\crefmultiformat{equation}{ (#2#1#3)}%
{ and~(#2#1#3)}{, (#2#1#3)}{ and~(#2#1#3)}

\DeclareFontFamily{U}{rsfs}{\skewchar\font127 }
\DeclareFontShape{U}{rsfs}{m}{n}{%
	<-6> rsfs5
	<6-8> rsfs7
	<8-> rsfs10
}{}
\usepackage[final]{microtype} %Addresses overfull \hbox in bibliography

\usepackage[utf8]{inputenc}  %Required by Inspire bib, but appears to not be necessary if compile with XeLaTeX.
\usepackage{textgreek}
\usepackage{tikz}
\usetikzlibrary{decorations.pathreplacing,angles,quotes,patterns}
\newcommand{\Amp}{\mathscr{M}}

\newcommand{\pt}{$p_T$}
\newcommand{\highpt}{high-\pt{}}

\DeclareMathOperator{\Tr}{Tr}

\newcommand{\fig}[1]{\cref{fig:#1}}
\newcommand{\eq}[1]{Eq.~\ref{eq:#1}}

\newcommand{\infinity}{\infty}
\newcommand{\sfracmu}{1/\mu}

\numberwithin{equation}{section}
\begin{document}

	\title{\bf Short path length corrections to Djordjevic-Gyulassy-Levai-Vitev energy loss}
	
	\author{Isobel Kolb\'{e}}
	\email[Corresponding author: ]{isobel.kolbe@cern.ch}
	\affiliation{Department of Physics, University of Cape Town, Private Bag X3, Rondebosch 7701, South Africa}
	\affiliation{National Institute for Theoretical Physics (NITheP), Western Cape, South Africa}
	
	\author{W.\ A.\ Horowitz} 
	\email{wa.horowitz@uct.ac.za}
	\affiliation{Department of Physics, University of Cape Town, Private Bag X3, Rondebosch 7701, South Africa}

	\begin{abstract}
		
		We compute the correction to the energy loss of a hard parton due to short separation distances between the creation of the particle and the in-medium scattering center that stimulates bremsstrahlung radiation, to first order in opacity.  In deriving the result we make full use of the large formation time assumption, which results in a significant reduction of the number of diagrams contributing to the small separation distance correction.  An asymptotic analysis of our small separation distance correction term finds that the correction dominates at large $ \sim 100 $ GeV parent parton energies; scales like $ L $ with the size of the system for small $ L $, but like $ L^0 $ at larger $ L $; and breaks color triviality.  An extensive numerical investigation of the correction term confirms the aforementioned analytic findings, reveals that the correction term does not go to zero for large $ L $, finds that the correction is sensitive to the mass of the parent parton, and shows a crucial dependence of the energy loss on a proper treatment of the physics of separation distances on the order of the Debye screening length.  However, upon examination, we have found the large formation time approximation to be invalid for much of the phase space of the emitted radiation, implying a need to investigate the sensitivity of jet quenching results from relaxing this approximation.  Our result constitutes an important step toward understanding partonic energy loss in small colliding systems.
	\end{abstract}


%---------------------------------------------------------------------------------------------------------------------------------------
%MAIN DOCUMENT
%---------------------------------------------------------------------------------------------------------------------------------------

\maketitle
\tableofcontents

	\section{Introduction}\label{sec:Introduction}
	
	Recent startling results from the Relativistic Heavy Ion Collider (RHIC) and the Large Hadron Collider (LHC) show that key signatures of quark-gluon plasma (QGP) formation are found in high-multiplicity p+p and p/d+A collision systems.  In particular, collective behavior \cite{Khachatryan:2015waa,Aad:2015gqa}, strangeness enhancement \cite{Abelev:2013haa,Adam:2015vsf}, and quarkonium suppression \cite{Du:2015wha,Adam:2016ohd} appear to be sensitive only to the measured multiplicity of the collision, and not to the size of the nuclear fireball as naively implied by the type of colliding particles.

	Jet quenching is another key observable of QGP formation \cite{Wiedemann:2009sh,Majumder:2010qh}, providing a unique femtoscope for probing the precise dynamics of the relevant degrees of freedom in this novel phase of nuclear matter.  Energy loss models \cite{Armesto:2005mz,Horowitz:2012cf,Burke:2013yra,Djordjevic:2014tka} based on perturbative quantum chromodynamics (pQCD) have had enormous qualitative success in describing the momentum dependence and angular distribution of the suppression of high-momentum ($\sim5-150$ GeV) single particle pions \cite{Adare:2013wop,Abelev:2014ypa} and charged hadrons \cite{CMS:2012aa,Abelev:2012hxa,Aad:2015wga,Shi:2018vys} from primordial hard light flavors, gluons, and electrons \cite{Abelev:2006db,Sakai:2013ata,Adare:2015hla}, as well as $D$ \cite{Adam:2015sza} and non-prompt $J/\psi$ mesons \cite{Chatrchyan:2012np} from open heavy flavor decays at mid rapidity in A+A systems from $\surd s = $ 0.2 ATeV to 5.02 ATeV.
	
	Early experimental analysis showed a tantalizing correlation between centrality and suppression of jets in p/d+A collisions at RHIC and LHC \cite{Adare:2015gla,ATLAS:2014cpa}, but later measurements have revealed that the experimental determination of jet quenching in small systems is fraught with difficulty \cite{Adam:2014qja}. These early results, along with newer measurements \cite{Adam:2016jfp} and sure-to-come future observations, call for quantitative theoretical predictions for jet tomography in small colliding systems.
	
	There are two major complications to comparing theoretical predictions to experimental measurements in small colliding systems.  First, phenomenologically, there is an inherent bias between rare high-multiplicity events and the rare collisions initially populated with one or more high transverse momentum (\highpt{}) particles \cite{Adam:2014qja,Armesto:2015kwa}. 
	Second, theoretically, derivations of energy loss based on pQCD use simplifying assumptions \cite{Armesto:2011ht} that make them inapplicable to a small brick of QGP.  
	
	The first complication makes it difficult to properly normalize the usual observable adopted in tomographic studies, the nuclear modification factor $R_{AB}$.  $R_{AB}$ is the ratio of a spectrum in A+B collisions to the same particle spectrum in p+p collisions suitably normalized such that $R_{AB}=1$ for particles unaffected by the presence of a QGP.  Because of the aforementioned bias, properly normalizing $R_{AB}$ in high-multiplicity p+p and p/d+A events is problematic.  One solution may be to divide the spectrum of interest by a known unaffected electroweak spectrum with the same event selection criteria, forming a $\gamma_{AB}$, $W_{AB}$, or $Z_{AB}$.  Another is to use a different centrality estimator \cite{Adam:2014qja}.
	
	The work of this article was motivated by the second complication. Although predictions of jet energy loss in small colliding systems have been put forward \cite{Zhang:2013oca,Park:2016jap}, they consistently over predict the observed suppression.  Such small system energy loss predictions utilize energy loss derivations that are derived for central and semi-central nucleus-nucleus collisions and it is therefore difficult to interpret the resultant discrepancy between theory and data as the absence of a hot thermal medium. In the usual DGLV (Djordjevic, Gyulassy, Levai, and Vitev) opacity expansion \cite{Gyulassy:2000er,Djordjevic:2003zk} for instance, the energy loss derivation assumes a large separation distance $\Delta z\equiv z_1-z_0\sim\lambda_{mfp}\gg1/\mu$ between the initial production position $z_0$ of the hard parent parton and the position $z_1$ where it scatters off a QGP medium quasiparticle.  This large separation allows one to 1) safely assume a factorization between the hard production process and the interaction of a nearly on-shell parton with a well-defined scattering center and 2) neglect several terms in the energy loss derivation.  The mean free path of the \highpt{} particle is $\lambda_{mfp} = 1/\rho\sigma\sim1-2$ fm while the Debye screening length in an infinite, static thermal QGP of temperature $T\sim350$ MeV is $\mu^{-1}=(gT)^{-1}\sim0.4$ fm, as derived from thermal field theory \cite{Wicks:2005gt}.  In the collision of p+p or p/d+A, one expects a system of radius $\lesssim2$ fm.  Therefore, for these small colliding systems, most \highpt{} particles have a separation distance between production and scattering that is not particularly large compared to the Debye screening length.

	In this article we modify the usual DGLV approach (see Sec.~\ref{sec:SetupAndCalculation} for details) by removing the second implication of the large separation distance assumption and retaining terms that were previously suppressed under the large separation distance assumption: we derive a generalization of the $N=1$ in opacity\footnote{In the reaction operator approach first put forward by GLV \cite{Gyulassy:2000er}, it was found that the induced gluon radiation of a hard parton, possibly undergoing multiple scatterings, in a dense medium, is dominated by the first order in opacity result.  That is, the gluon radiation is dominated by the $ N=1 $ contribution, where $ N $ refers to the number of scatterings that the hard parton or radiated gluon undergoes with the medium.} DGLV radiative energy loss result \cite{Djordjevic:2009cr,Djordjevic:2008iz} by including all previously neglected terms assumed small under the scale ordering $\Delta z \gg 1/\mu$; see Fig.~\ref{fig:Scattering}.  Note that the inclusion of smaller separation distances does not affect the scale of the Debye screening length in relation to the mean free path, which is to say that the Gyulassy-Wang model \cite{Gyulassy:1993hr}, used to model the target scatterer (see Eq.~\ref{eq:GWPotential}), remains valid.  Since the formation time for a \highpt{} particle goes as $\tau_f\sim1/p_T\lesssim 1/\mu$, our derivation is fully justified for $\Delta z\gtrsim 1/\mu\sim0.4$ fm for $p_T\gtrsim$ few GeV.  To the extent that factorization, near-on-shellness, and the Gyulassy-Wang model for scattering centers are good approximations even when $\Delta z\lesssim 1/\mu$, we have thus derived the \emph{all} separation distances generalization of $N=1$ in opacity DGLV energy loss.  Note also that the present short separation distance correction is an additional incorporation of finite size effects, over and above the effects that are due to producing the parent parton at finite time (as opposed to the infinite past), as computed by \cite{Djordjevic:2009cr,Djordjevic:2008iz} 
	
	\begin{figure}[!t]
		
		\definecolor{ColomBlue}{RGB}{188,232,255}
		\centering
		\begin{tikzpicture}
		% Big box
		\draw [fill=ColomBlue, ultra thick] (0,7) rectangle (4,5);
		\draw [outer color=orange!70!black,inner color=white,minimum width=3cm,draw=none] (0.5,6) circle [radius=0.3];
		\draw [outer color=orange!70!black,inner color=white,minimum width=3cm,draw=none] (2,5.5) circle [radius=0.3];
		\draw [outer color=orange!70!black,inner color=white,minimum width=3cm,draw=none] (3.5,6.5) circle [radius=0.3];
		
		\draw [dashed,ultra thick](1,7) -- (1,5);
		
		%Top label
		\node[scale=1.3] at (2,8) {$\frac{1}{\mu}\ll\Delta z\sim\lambda_{mfp}\ll L$};
		
		%Top curly brace
		
		\draw[decoration={brace,amplitude=10pt},decorate,thick] (0,7.3) -- (4,7.3);
		
		%Bottom label
		\node[scale=1.3] at (1,4) {$\frac{1}{\mu}\ll\lambda_{mfp}$};
		
		%Bottom curly brace
		
		\draw[decoration={brace,amplitude=7pt,mirror},decorate,thick] (0,4.8) -- (1,4.8);
		
		\end{tikzpicture}
		\vspace{-.1in}
		\caption{The usual DGLV setup (full box) compared to the setup used in this article (left of the dashed line), showing a static QGP brick of length $ L $, containing arbitrarily distributed scattering centers (orange balls).  Left of the dashed line, no statement is made regarding $ \Delta z $, the distance between hard production and first scattering, allowing for an application to small systems where $L \sim 1/\mu$.
			\label{fig:Scattering}}
	\end{figure}
	
	Phenomenological energy loss models perform an average over the position(s) at which scattering(s) occur in the given distance that a parton travels in medium, $L$.  Therefore, even though no previous energy loss derivation correctly treated the region $\Delta z\lesssim1/\mu$, all energy loss models nonetheless used the derived energy loss formulae in this region.  One might have hoped that the use of these formulae in regions where they are invalid (when $ \Delta z\lesssim 1/\mu$) could be justified either by an argument that the small separation distance corrections are small, or by an explicit {\it a posteriori} calculation.  What we find from the calculation presented in this work is that the short distance correction can be very large, especially as the momentum of the parent parton becomes large.  Worse still, the physics of the early times $ \tau\leq\tau_0 $ is not at all clear.
	
	The simplified picture employed by DGLV, in which a hard parton traverses a brick of QGP modeled by a collection of thermalized quasi-particles, does not necessarily explicitly incorporate all the phenomenological aspects of partonic energy loss in a heavy ion collision.  Particular concerns for energy loss phenomenology include the factorization of the hard parton's production in the presence of large fields from its propagation through the fireball, the effect of a boundary on the shape of the Debye screened scattering centers, and the time required for the medium to thermalize and form scattering centers for the hard parton to interact with.  In order to investigate the importance of this lack of knowledge, we explore various distributions of scattering centers.  We find that the original DGLV is insensitive to the details of the physics at small separation distances $ \Delta z\lesssim\sfracmu $.  This insensitivity is due to a delicate cancellation of interfering terms, a cancellation beyond formation time effects.  On the other hand, the cancellation is not quite so precise for the correction term, which leads to a significant dependence of the correction term on the details of the short distance physics. 
	
	The DGLV formalism (see Sec.~\ref{sec:SetupAndCalculation} for more details) includes the assumption that the formation time of the radiated gluon is much larger than the Debye screening length. We will see that this ``large formation time'' assumption will play a crucial role in the derivation of the small separation distance correction term, resulting in a major reduction in the number of diagrams that contribute to the small separation distance correction.  We will further show that, not only does the formation time set an important scale for the understanding of the early time physics of the correction term, but also that the large formation time assumption is invalid for much of the relevant gluon emission phase space.  Previous work has demonstrated the extreme sensitivity of \emph{all} energy loss calculations to the collinear approximation \cite{Horowitz:2009eb,Armesto:2011ht}, and therefore the need to move beyond its use in all energy loss models.  However, we emphasize that the sensitivity we find from the large formation time approximation is both new and different from the sensitivity to the collinear approximation.  As such, all current jet quenching models that include radiative energy loss based on pQCD must individually assess their sensitivity to the large formation time and large system size approximations when making quantitative comparisons with data.

	\section{Setup and calculation}\label{sec:SetupAndCalculation}

	In this article we use precisely the setup of the DGLV calculation \cite{Djordjevic:2003zk}, presenting here only an outline of the setup and derivation of the correction term, with more details in \cite{Kolbe:2015uka}.  For clarity, we treat the  \highpt{} eikonal parton produced at an initial point $(t_0,z_0,\textbf{x}_0)$ inside a finite QGP brick, where we have used \textbf{p} to mean transverse 2D vectors, $\vec{\textbf{p}}=(p_z,\textbf{p})$ for	3D vectors and $p=(p^0,\vec{\textbf{p}})=[p^0+p^z,p^0-p^z,\textbf{p}]$ for four vectors in Minkowski and light cone coordinates, respectively.  As in the DGLV calculation, we consider the $n$th target to be described by a Gyulassy-Wang Debye screened potential \cite{Gyulassy:1993hr} with Fourier and color structure given by			
	\begin{align}\label{eq:GWPotential}
	V_n
	&= 2\pi\delta(q^0)\frac{4\pi\alpha_s}{\vec{q}_n^2+\mu^2}e^{-i\vec{q}_n\cdot \vec{x}_n}T_{a_n}(R)\otimes T_{a_n}(n),
	\end{align}
	where the color exchanges are handled using the applicable $S\!U(N_c)$ generator, $T_a(n)$ in the $d_n$ dimensional representation of the target, or $T_a(R)$ in the $d_R$ dimensional representation of the \highpt{} parent parton.
	
\begin{figure*}[!t]
		\centering
		\includegraphics[scale=0.6]{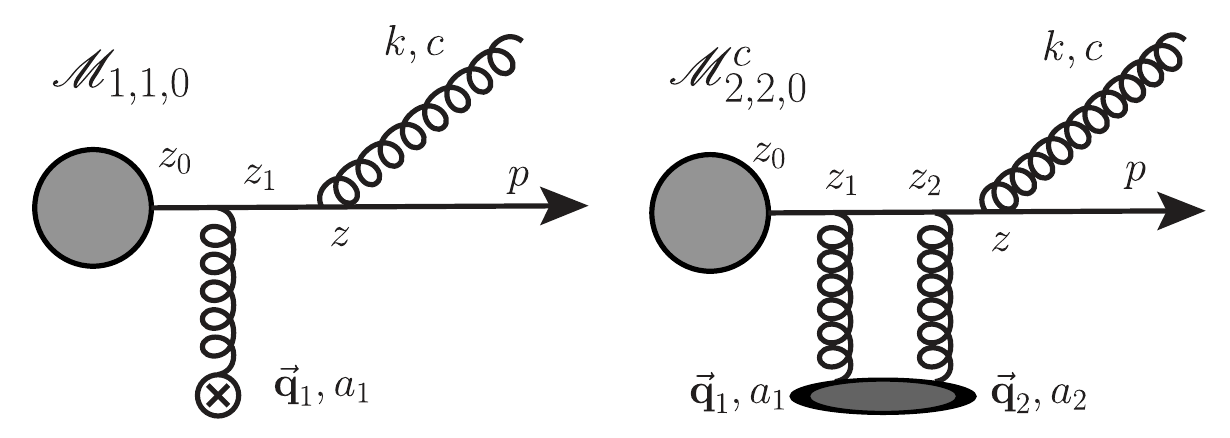}
		\vspace{-.1in}
		\caption{\label{fig:diagrams}Following the diagrammatic numbering in \cite{Djordjevic:2003zk}, $\Amp_{1,1,0}$ (left-hand panel) and $\Amp^c_{2,2,0}$ (right-hand panel) are the only two diagrams that have non-zero small separation distance corrections in the large formation time limit.  $\Amp^c_{2,2,0}$ is the double Born contact diagram, corresponding to the second term in the Dyson series in which two gluons are exchanged with the single scattering center.}
	\end{figure*}

	In light cone coordinates the four-momenta of the emitted gluon, the final \highpt{} parton, and that exchanged with the medium Debye quasiparticle are, respectively,
	\begin{align}\label{Eqn:momenta}
	k	&=\bigg[xP^+, \frac{m_g^2+\textbf{k}^2}{xP^+},\textbf{k}\bigg],\nonumber\\
	p	&=\bigg[(1-x)P^+,\frac{M^2+\textbf{k}^2}{(1-x)P^+},\textbf{q}-\textbf{k}\bigg],\nonumber\\
	q 	&=[q^+,q^-,\textbf{q}],
	\end{align} 
	where the initially produced \highpt{} particle of mass $M$ has large momentum $E^+=P^+=2E$ and negligible other momentum components.  Notice that we include the QCD analogue of the Ter-Mikayelian plasmon effect \cite{Djordjevic:2003be}, a color dielectric modification of the gluon dispersion relation, with an effective emitted gluon mass $m_g$ \cite{Djordjevic:2003zk,Wicks:2005gt}.  See \fig{diagrams} for a visualization of the relevant momenta.
	
	Following \cite{Djordjevic:2003zk} we define 
	$\omega\approx xE^+/2 = xP^+/2$, from which a shorthand for energy ratios will prove useful notationally:   
	$\omega_0\equiv \textbf{k}^2/2\omega$, 
	$\omega_i\equiv (\textbf{k}-\textbf{q}_i)^2/2\omega$, 
	$\omega_{(ij)}\equiv (\textbf{k}-\textbf{q}_i-\textbf{q}_j)^2/2\omega$, and 
	$\tilde{\omega}_m\equiv \big(m_g^2+M^2x^2\big)/2\omega$.  
	
	We will also make the following assumptions: 1) the eikonal, or high energy, approximation, in which $E^+$ is the largest energy scale of the problem; 2) the soft\footnote{The validity of the soft gluon approximation within the DGLV formalism has recently been confirmed explicitly \cite{Blagojevic:2018nve}. } (radiation) approximation $x \ll 1$; 3) collinearity, $k^+\gg k^-$; 4) that the impact parameter varies over a large transverse area; and, most crucially for the present article, 5) the large formation time assumption $\omega_i \ll \mu_i$, where $\mu_i^2\equiv\mu^2+\textbf{q}_i^2$.
	
	Note that the above approximations, in addition to allowing us to systematically drop terms that are small, permit us to 1) (eikonal) ignore the spin of the \highpt{} parton; 2) (soft) assume the source current for the parent parton varies slowly with momentum $J(p-q+k)\approx J(p+k)\approx J(p)$; 3) (collinearity) complete a separation of energy scales,
	\begin{equation}
	E^+\gg k^+\gg k^- \equiv \omega_0 \sim\omega_{(i\ldots j)}\gg\frac{(\textbf{p}+\textbf{k})^2}{P^+};
	\end{equation}
	and 4) (large area) take the ensemble average over the phase factors, which become $\langle e^{-i(\textbf{q}-\textbf{q}')\cdot \textbf{b}}\rangle=\frac{(2\pi)^2}{A_\perp}\delta^2(\textbf{q}-\textbf{q}').$
	
	In the original DGLV calculations \cite{Djordjevic:2003zk}, the large formation time approximation played only a minor role.  However, when considering small separation distances between the scattering centers, the large formation time assumption naturally increases in importance.  
	
	With the above approximations, we reevaluated the 12 diagrams contributing to the $N=1$ in opacity energy loss amplitude \cite{Djordjevic:2003zk} without the additional simplification of the large separation distance $\Delta z \gg 1/\mu$ assumption.  
	
	In the original evaluation of the 12 diagrams contributing to the $N=1$ in opacity energy loss derivation, the large separation distance approximation $\Delta z\gg1/\mu$ allowed for the neglect of terms proportional to $\exp(-\mu \Delta z)$.  In our reevaluation of these 12 diagrams we retained all terms proportional to $\exp(-\mu \Delta z)$. However, we found an enormous simplification due to the large formation time approximation $\omega_i\ll\mu_i$: all but two of the 12 diagrams' 18 new small separation distance correction pole contributions are suppressed under the large formation time assumption.  We show the two diagrams\footnote{The diagrammatic numbering employed here is chosen for ease of reference when comparing to previous GLV calculations.  The subscripts pertain to factorizations in the reaction operator approach \cite{Gyulassy:2000er} and have no significance here beyond a convenient naming convention.} $\Amp_{1,1,0}$ and $\Amp^c_{2,2,0}$ with non-zero contributions at the amplitude level in the large formation time approximation in \fig{diagrams}.  See \cite{Kolbe:2015uka} for the computation of all 12 relevant diagrams.  One can see from \fig{diagrams} that the class of diagrams that contribute to the short distance correction is that for which the radiated gluon is emitted subsequent to the parent parton scattering off the medium.

	The reason for the contributing class of diagrams being those for which the scattering occurs prior to the emission of the gluon is the competition between relaxing the large distance approximation $\Delta z \gg 1/\mu$ and keeping the large formation time approximation, $\tau_{form} = xE/\mathbf{k}^2 \gg 1/\mu$.  For a diagram to contribute to the small separation distance correction, we require $\Delta z \lesssim 1/\mu$.  However, if the gluon is emitted at $\Delta z \lesssim \tau_{form}$, the large formation time approximation dictates that the gluon is not formed before the parent parton encounters a scattering center. The scattering center cannot therefore resolve the gluon independently from the parent parton, and these diagrams' contributions are suppressed.

	\begin{widetext}
	The full result for these two amplitudes under our approximation scheme is then
	
		\begin{alignat}{2}
		&\Amp_{1,1,0}	
		&&\approx-J(p)e^{ipx_0}2gT_{a_1}ca_1\int\frac{d^2\textbf{q}_1}{(2\pi)^2}v(0,\textbf{q}_1)
		e^{-i\textbf{q}_1\cdot\textbf{b}_1}\times\frac{\textbf{k}\cdot\boldsymbol\epsilon}{\textbf{k}^2+m_g^2+x^2M^2}
		\bigg[e^{i(\omega_0+\tilde{\omega}_m)(z_1-z_0)}
		-\frac{1}{2}e^{-\mu_1(z_1-z_0)}\bigg]\\
		&\Amp_{2,2,0}^c		
		&&\approx  J(p)e^{i(p+k)x_0}\int\frac{d^2\textbf{q}_1}{(2\pi)^2}\int\frac{d^2\textbf{q}_2}{(2\pi)^2}
		e^{-i(\textbf{q}_1+\textbf{q}_2)\cdot\textbf{b}_1}
		\times ig T_{a_2}T_{a_1}ca_2a_1v(0,\textbf{q}_1)v(0,\textbf{q}_1)
		\times\frac{\textbf{k}\cdot\boldsymbol\epsilon}{\textbf{k}^2+m_g^2+x^2M^2}\nonumber\\
		&	&&\times \bigg[e^{i(\omega_0+\tilde{\omega}_m)(z_1-z_0)}+e^{-\mu_1(z_1-z_0)}
		\bigg(1-\frac{\mu_1e^{-\mu_2(z_1-z_0)}}{2(\mu_1+\mu_2)}\bigg)\bigg].
		\end{alignat}

	The double differential single inclusive gluon emission distribution is given by \cite{Djordjevic:2003zk}
	\begin{align}
	d^3N_g^{(1)}d^3N_J	
	&=\frac{d^3\vec{\textbf{p}}}{(2\pi)^32p^0}\frac{d^3\vec{\textbf{k}}}{(2\pi)^32\omega}\times\bigg(\frac{1}{d_T}\Tr\langle\vert\Amp_{1}\vert^2\rangle+\frac{2}{d_T}\Re\Tr\langle\Amp^*_0\Amp_2\rangle\bigg),
	\end{align}
	from which the energy loss, given by the energy-weighted integral over the gluon emission distribution $\Delta E = E\int \! dx \, x dN_g/dx$, can be calculated from the amplitudes.  
	
	Our main analytic result is then the $N=1$ first order in opacity all separation distance generalization of the DGLV induced energy loss of a \highpt{} parton in a QGP:
	
	\begin{align}\label{eq:ELossFull}
	\Delta 
	& E_{ind}^{(1)} =\frac{C_R\alpha_sLE}{\pi\lambda_g}	\int dx\int \frac{d^2\textbf{q}_1}{\pi}\frac{\mu^2}{(\mu^2+\textbf{q}_1^2)^2}
	\int\frac{d^2\textbf{k}}{\pi} \int d\Delta z\;\rho(\Delta z) \nonumber\\ 
	& \times \Bigg[-\frac{2\big(1-\cos\big\{(\omega_1+\tilde{\omega}_m)\Delta z\big\}\big)}
	{(\textbf{k}-\textbf{q}_1)^2+m_g^2+x^2M^2}
	\bigg(\frac{(\textbf{k}-\textbf{q}_1)\cdot\textbf{k}}{\textbf{k}^2+m_g^2+x^2M^2}
	-\frac{(\textbf{k}-\textbf{q}_1)^2}{(\textbf{k}-\textbf{q}_1)^2+m_g^2+x^2M^2}\bigg)\nonumber\\
	&	+\frac{1}{2}e^{-\mu_1\Delta z}\Bigg\{ \bigg(\frac{\textbf{k}}{\textbf{k}^2+m_g^2+x^2M^2}\bigg)^2\bigg(1-\frac{2 C_R}{C_A}\bigg)
	\bigg(1-\cos\{(\omega_0+\tilde{\omega}_m)\Delta z\}\bigg)\nonumber\\
	&	+	\frac{\textbf{k}\cdot (\textbf{k}-\textbf{q}_1)}
	{\big(\textbf{k}^2+m_g^2+x^2M^2\big)\big((\textbf{k}-\textbf{q}_1)^2+m_g^2+x^2M^2\big)}\big(\cos\{(\omega_0+\tilde{\omega}_m)\Delta z\}-\cos\{(\omega_0-\omega_1)\Delta z\}\big)\Bigg\}\Bigg].
	\end{align}
	\end{widetext}
	
	The second line in Eq.~\ref{eq:ELossFull} (along with the prefactor in the first line) is the original DGLV result (herein after ``the DGLV'' term).  The last two lines are the small separation distance correction (herein after ``the correction'' or ``the small separation distance correction'').  It is natural to define $\Delta E_{ind}^{(1)} \equiv \Delta E_\mathrm{DGLV}^{(1)} + \Delta E_\mathrm{corr}^{(1)}$.  In what follows we will refer to the full DGLV + correction in Eq.~\ref{eq:ELossFull} as the ``all separation distance'' result.  The correction term has the properties we expect: 1) the correction goes to zero as the separation distance becomes large, $\Delta z\rightarrow\infty$ (or, equivalently, as the Debye screening length goes to 0, i.e.,\ $\mu\rightarrow \infty$) and 2) the correction term vanishes as the separation distance vanishes, $\Delta z\rightarrow 0$, due to the destructive interference of the Landau-Pomeranchuk-Migdal (LPM) effect.
	
	An immediate surprise is the breaking of color triviality, whereby the energy loss depends only on the representation of the parent parton via the Casimir $ C_R $ in the representation $ R $ of the parton - there is no dependence upon a preferred representation for a color trivial result. Although color triviality is seen to all orders in opacity in the large separation distance approximation \cite{Gyulassy:2000er}, the color triviality breaking in the small separation distance correction comes from the term proportional to $2C_R/C_A$. We will investigate the effect of this term numerically in Sec.~\ref{sec:ColorTriviality}.

	\section{Numerical and asymptotic analyses}\label{sec:NumericalAspectsAndAsymptoticAnalysis}
	
	\begin{figure*}
		\centering
		\includegraphics[scale=0.6]{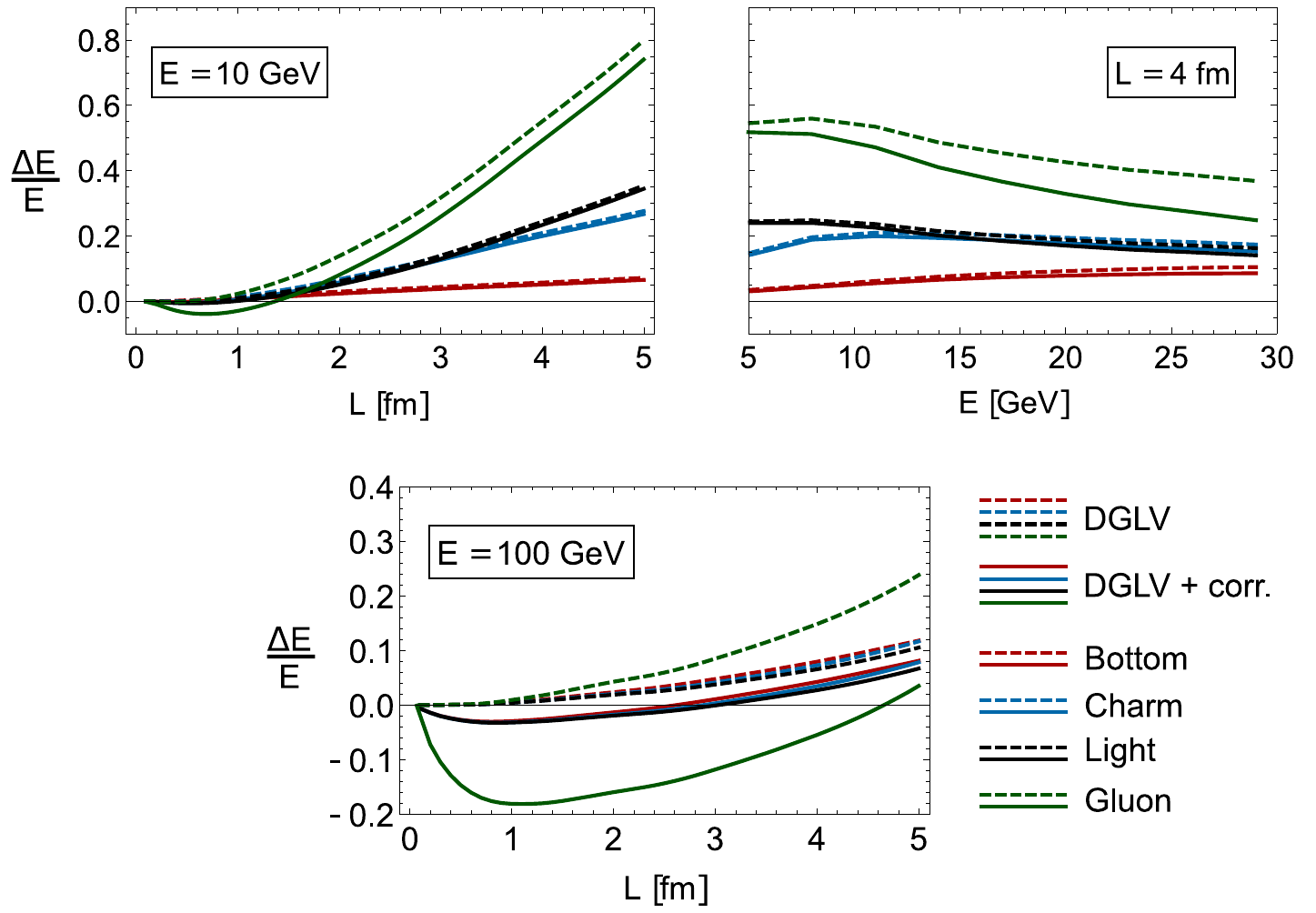}
		\caption{(Color online) (Grayscale coloring guide: green, black, blue, red from top to bottom on the right hand side of the top left plot and on the left hand side of the top right plot; dashed green, dashed red, dashed blue, dashed black, solid red, solid blue, solid black, solid green from top to bottom on the right hand side of the bottom plot) Fractional energy loss of bottom (red), charm (blue), and light quarks (black), as well as gluons (green) in a QGP with $\mu = 0.5$ GeV and $\lambda_{mfp} = 1$ fm for (top left) fixed energy $E = 10$ GeV, (top right) fixed path length $L = 4$ fm, and (bottom) fixed energy $ E = 100 $ GeV.  Here, DGLV curves (dashed) are computed from the original $N=1$ in opacity large separation distance DGLV formula while DGLV + corr. curves (solid) are from our all separation distance generalization of the $N=1$ DGLV result, \protect\eq{ELossFull}.\label{Fig:DGLVvsDGLVHKFtoEandtoLPlot}}
	\end{figure*}	
	
	Fig.~\ref{Fig:DGLVvsDGLVHKFtoEandtoLPlot} is produced by computing the $ \Delta z $ integral in \eq{ELossFull} analytically before computing all other integrals numerically (we will refer to this process as the ``numerical investigation'').
	The numerical results use the same values as \cite{Djordjevic:2003zk}: $\mu = 0.5$ GeV, $\lambda_{mfp}=1$ fm, $C_F=4/3$, $C_A=3$, $\alpha_s=0.3$, $m_{c}=1.3$ GeV, $m_{b}=4.75$ GeV, and $ m_{q}=\mu/2 $ (light quark mass) \cite{Wicks:2005gt}.  The QCD analogue of the Ter-Mikayelian plasmon effect is taken into account by setting $m_{g}=\mu/\! \sqrt{2}$ \cite{Djordjevic:2003be}.  As in \cite{Wicks:2005gt}, kinematic upper limits are used for the momentum integrals such that $0 \leq k\leq 2x(1-x)E$ and $0\leq q \leq \sqrt{3 E \mu}$.  This choice of $k_{max}$ guarantees that the final momentum of the parent parton is collinear to the initial momentum of the parent parton and that the momentum of the emitted gluon is collinear to the momentum of the parent parton. The fraction of momentum carried away by the radiated gluon $x$ is integrated over from $0$ to $1$.  The distribution of scattering centers $ \rho(\Delta z) $, although originally assumed to be exponential in \cite{Djordjevic:2003zk}\footnote{Choosing an exponential distribution for $ \rho(\Delta z)$ was done in order to account for the rapidly expanding medium as well as to allow for clever manipulations leading to a deeper understanding of the asymptotic behavior of the formula, since the exponential form relates well to the cosines in the energy loss formula.}, is assumed (in Fig.~\ref{Fig:DGLVvsDGLVHKFtoEandtoLPlot}) to have the form of a unit step function, since an exponential distribution biases toward short separation distance scattering, lending potentially excessive weight to contributions from short separation distance terms.  
	%The choice of a step function distribution reduces the effect of the correction terms by $\sim 10\%$ at low ($\sim 10$ GeV) parton energies and $\sim 50\%$ at higher ($\sim 100$ GeV) energies as compared to results using an exponential distribution.  We will elaborate further on the choice of scattering center distribution function later in this section.

	In the top left-hand panel of Fig.~\ref{Fig:DGLVvsDGLVHKFtoEandtoLPlot} we plot the fractional energy loss of various parent partons of energy $E=10$ GeV for path lengths up to 5 fm.  One sees that the correction has a non-negligible effect even for large path lengths.  Although initially unanticipated, the fact that the correction is substantial even for  $L	 \gtrsim 3 $ fm (perhaps most easily seen for gluons, but the relative size of the correction is meaningful even for the quarks), is due to the integration over all separation distances between the production point and the scattering position; even for large path lengths, some of the interaction distances between the parent parton and the target occur at separation distances that are small compared to the Debye screening scale.  However, as expected, the relative size of the correction term and the leading DGLV result diminishes at fixed energy as the path length grows.
	
	In the top right-hand panel of Fig.~\ref{Fig:DGLVvsDGLVHKFtoEandtoLPlot} we show the fractional energy loss of parent partons of varying energy, propagating through an $ L=4 $ fm long static QGP brick.  Notice first that the small separation distance correction term is generally an energy \emph{gain} due to the sign of the color triviality breaking term and, second, that the size of the correction relative to the large separation distance DGLV result grows with energy.  
	
	In the bottom panel of Fig.~\ref{Fig:DGLVvsDGLVHKFtoEandtoLPlot} we show the fractional energy loss of parent partons of energy $ E=100 $ GeV. 
	In this $ E=100$ GeV plot one sees that the short separation distance correction term dominates over the DGLV term out to distances of $ L\sim3-5 $ fm $ \gg \sfracmu $.  We investigate this surprisingly persistent domination further in Sec.~\ref{sec:EnergyDependence}.  One further observes that the color factor in the correction term plays a crucial role, since the gluon energy loss is dramatically different from quark energy loss, especially at high energies (investigated further in Sec.~\ref{sec:ColorTriviality}).

	\subsection{Color triviality}\label{sec:ColorTriviality}
	
	The color triviality breaking term in the small separation distance correction means the correction for gluons can be an order of magnitude larger than for quarks.
	To see this difference, consider the first line of the correction term in Eq.~\ref{eq:ELossFull} which contains a term that carries the factor $ \left(1-\frac{2C_R}{C_A}\right) $.  For gluons, $ C_R=C_A=3 $ while for quarks $ C_R=C_F=\sfrac{4}{3} $, giving a factor of $ -1 $ for gluons and a factor of $ \sfrac{1}{9} $ for quarks. This factor of ten difference means that, although the gluons have an effective mass (as a result of the way in which the QCD analogue of the Ter-Mikayelian plasmon effect was taken into account) that is only marginally larger than the plasmon mass of the light quarks, the gluons will not necessarily obey the same mass ordering as the quarks.
	
\begin{figure*}
		\centering
		\includegraphics[scale=0.8]{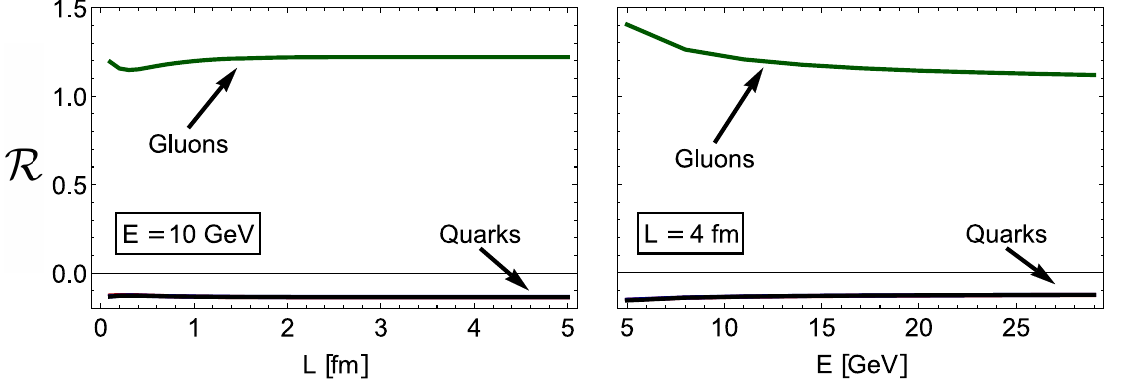}
		\caption{The ratio $ \mathcal{R} $ of the color triviality breaking and the color trivial parts of the correction term in Eq.~\ref{eq:ELossFull}, for quarks ($ C_F $) and gluons ($ C_A $), as a function of the length $ L $ of the brick for parent partons with $ E=10 $ GeV (left-hand panel), and as a function of the energy $ E $ of a parent parton moving through a brick of length $ L=4 $ fm (right-hand panel).\label{fig:HKtoEColorRatioPlotPaper} }
	\end{figure*}

	To illustrate this effect, we have plotted in Fig.~\ref{fig:HKtoEColorRatioPlotPaper} the ratio $ \mathcal{R} $ of the color trivial and color nontrivial terms of the correction term; i.e., we have divided the color triviality breaking part of the correction term, proportional to $ \left(1-\frac{2C_R}{C_A}\right)  $, by the color trivial part of the correction term. We show this ratio both as a function of the length $ L $ of the brick (left-hand panel) and as a function of the parent parton energy $ E $ (right-hand panel).  Fig.~\ref{fig:HKtoEColorRatioPlotPaper} clearly shows the order of magnitude difference between the color trivial and color nontrivial parts of the correction term for parent partons in the fundamental and adjoint representations, the difference in sign of the correction for quarks and gluons, and the persistence of the difference in magnitude of the correction as a function of both $ L $ and $ E $.

	\subsection{Energy dependence and asymptotic analysis}\label{sec:EnergyDependence}
	
	A striking feature of the plot in the bottom panel of Fig.~\ref{Fig:DGLVvsDGLVHKFtoEandtoLPlot} is the dominance of the small separation distance correction term at high energies.  We see in Fig.~\ref{Fig:DGLVvsDGLVHKFtoEandtoLPlot}, by comparing the top left-hand panel to the bottom panel, a dominance of the correction term at $ E=100 $ GeV, leading to an energy gain, even out to systems with sizes of $ L\sim 3 $ fm for quarks and $L \sim 5 $ fm for gluons. In order to better understand this dominance of the correction term at large energies, one may perform an asymptotic analysis.	Recall that $\Delta E_{ind}^{(1)} \equiv \Delta E_\mathrm{DGLV}^{(1)} + \Delta E_\mathrm{corr}^{(1)}$, where $\Delta E_{ind}^{(1)}$ is given by \eq{ELossFull}.  Starting with the correction term $\Delta E_\mathrm{corr}^{(1)}$ and following \cite{Gyulassy:2000er}, we take all thermal and quark masses to zero, and analytically evaluate the integral over the scattering separation distance $\Delta z$ (using an exponential distribution for analytic simplicity and to connect with the known analytic results of \cite{Gyulassy:2000er}).  Then, we remove the kinematic bound on the momentum kick from the medium $q_{max}\rightarrow\infinity$, shift the momentum integral, analytically evaluate the angular integrals in momentum space, and perform the integrals over $k$ and $q$.  The result is
	
	\begin{equation}
	\label{eq:asympk}
	\Delta E_\mathrm{corr}^{(1)} 
	= \frac{C_R \alpha_s}{2\pi}\frac{L}{\lambda_g}\bigg(-\frac{2 C_R}{C_A}\bigg)\frac{12}{2+\mu L}E 
	\int_{0}^{1}dx \log\bigg(\frac{L\,k_{max}}{2+\mu L}\bigg).
	\end{equation}
	
	Taking, for simplicity, $k_{max}=2xE$ we find
	\begin{equation}
	\label{eq:AsympNLO}
	\Delta E_\mathrm{corr}^{(1)} = \frac{C_R \alpha_s}{2\pi}\frac{L}{\lambda_g}\bigg(-\frac{2 C_R}{C_A}\bigg)\frac{1}{2+\mu L}E \, \log\bigg(\frac{2 E L}{2+\mu L}\bigg)
	\end{equation}
	in the limit of large energy $E$.

	The equivalent asymptotic expression for the large separation distance leading massless DGLV result was derived in \cite{Gyulassy:2000er}.  The result, with $k_{max}\rightarrow\infinity$, is
	\begin{equation}\label{eq:AsympLO}
	\Delta E^{(1)}_\mathrm{DGLV}=\frac{C_R\alpha_s}{4}\frac{L^2\mu^2}{\lambda_g}\log\frac{E}{\mu}.
	\end{equation}
	
	There are several important features of Eqs.~\ref{eq:AsympNLO} and \ref{eq:AsympLO} to note.  First, the terms not proportional to the color triviality breaking $2C_R/C_A$ factor in \eq{AsympNLO} cancel at this level of approximation since $ k_{max}\gg q_{max} $, and the correction is purely an \emph{energy gain}.  Second, the correction term is log divergent in the upper bound of the perpendicular momentum of the emitted gluon $k_{max}$, whereas the large separation distance DGLV term is finite for infinite $k_{max}$.  Third, the correction term is linear in $L$ at small $ L $ and independent of $ L $ at large $ L $, while the DGLV term is proportional to the usual $L^2$.  Fourth, the asymptotic correction term breaks color triviality as its magnitude is proportional to $L/\lambda_R$, where $\lambda_R$ is the mean free path of the parent parton (whether quark or gluon), instead of proportional to $L/\lambda_g$, where $\lambda_g$ is the mean free path for gluons.
	
	Most important, cancellation between the contributions to the large separation distance DGLV result leads to an energy loss that grows only logarithmically with energy $E$.  The small separation distance correction piece does not suffer from a similar interference and grows \emph{linearly} with $E$ (with logarithmic in $ E $ correction).  It is precisely this linear in $E$ behavior compared to the log $E$ of the large separation distance DGLV term that leads to the correction term dominating over the leading term at higher energies.  The subtle cancellation between terms in the DGLV term, and the absence of such a cancellation in the correction term is discussed in more detail in Sec.~\ref{sec:EarlyTimeSensitivity}.
	
	The fact that the short separation distance ``correction'' term can dominate over the leading large separation distance DGLV result even out to path lengths $L\sim4\mu$ when not relaxing the large formation time assumption (effects that should tend to zero under the large formation time assumption), suggests that the large formation time assumption is invalid in the DGLV approach.  The dependence of the energy loss on the large formation time assumption is explored further in Sec.~\ref{sec:EarlyTimeSensitivity} as well.

	\subsection{Mass ordering and the large formation time assumption}\label{sec:MassOrdering}

	In Fig.~\ref{Fig:DGLVvsDGLVHKFtoEandtoLPlot}, the all separation distance energy loss can be seen to be mass ordered\footnote{Note that this mass ordering does not hold for the gluons, even though they take on an effective plasmon mass.  This is due to the color factor in the correction term; see Sec.~\ref{sec:ColorTriviality}.}.  The mass ordering of the large separation distance relative energy loss was found in \cite{Djordjevic:2003zk}, where the explanation was that the effect of increasing the mass of the parent parton was to reduce the relevance of the gluon formation time factor. 
	The formation time physics of the large separation distance DGLV result is encoded in the cosine terms of Eq.~\ref{eq:ELossFull} and a similar dependence on gluon formation times is apparent in the small separation distance correction term. 
	
\begin{figure}
		\centering
		\includegraphics[scale=0.4]{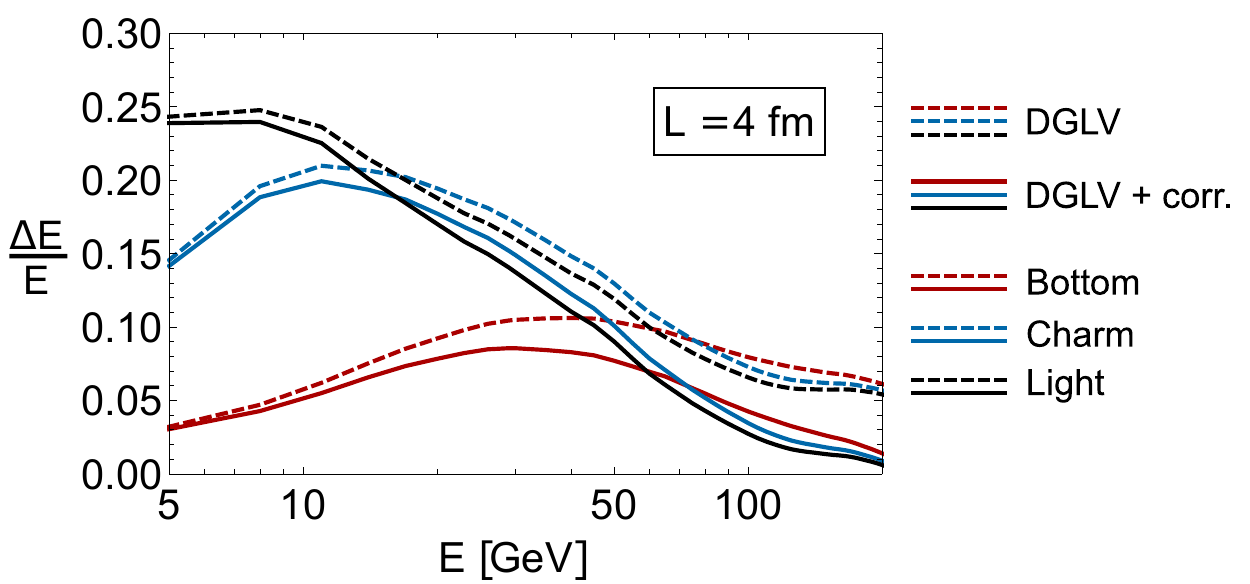}
		\caption{(Color online) (Grayscale coloring guide: top to bottom on the left hand side are black, blue, red, respectively) Fractional energy loss of bottom (red), charm (blue), and light (black) quarks in a GQP with $ \mu =0.5$ GeV and $ \lambda_{mfp}=1 $ fm for fixed path length $ L= 4$ fm.  The gluon result has not been plotted here because the color factor means that the gluon result does not obey the mass ordering discussed in this section.  Here DGLV curves (dashed) are computed from the original $ N=1 $ opacity large separation distance DGLV formula while DGLV + corr. curves (solid) are from our all separation distance generalization of the $ N=1 $ DGLV result, Eq.~\ref{eq:ELossFull}.  \label{fig:DGLVvsDGLVHKFtoElongPlot}}
	\end{figure}

	However, notice that the mass dependence of Eq.~\ref{eq:ELossFull}, is also apparent in the massive propagator.  The propagator masses lead straightforwardly to a  reduction of energy loss.  At low energies the propagator mass ordering dominates the energy loss, leading to higher mass partons losing less energy.  On the other hand, since the prefactors containing the propagators scale like $ 1/E^2 $ while the formation times scale like $ 1/E $, formation time physics dominates the mass dependence of the energy loss at high energies; formation times are shorter for more massive parent partons, leading to an enhancement towards incoherent energy loss.  
	We may thus understand the inversion of the mass ordering in the top right- hand panel of Fig.~\ref{Fig:DGLVvsDGLVHKFtoEandtoLPlot} (at $ E=10 $ GeV) which results from the massive propagator, to the ordering observed in the bottom panel of Fig.~\ref{Fig:DGLVvsDGLVHKFtoEandtoLPlot} (at $ E=100 $ GeV) where the mass ordering is dominated by formation time physics.  In order to see this more clearly, we present Fig.~\ref{fig:DGLVvsDGLVHKFtoElongPlot}, where it is clear that the mass ordering at low energy is the opposite to that at high energy.  Note also that the inversion of the mass ordering holds for the original DGLV result. 
	
	Nevertheless, despite the weak dependence of the mass ordering of the relative energy loss on the gluon formation time at low energies, recall the crucial role that the large formation time approximation $ \omega_i\ll \mu_i $ plays in the derivation of the small separation distance correction.  Traditionally, the large formation time assumption is considered a restatement of the collinear radiation approximation, but it is already known that the collinear assumption is problematic \cite{Armesto:2011ht}:
	it was shown in \cite{Horowitz:2009eb} that a significant fraction of the gluon radiation from $N=1$ large separation distance DGLV is \emph{not} emitted collinearly, despite the use of the collinear approximation $k^+\gg k^-$ in the derivation of the result.  One may understand this breakdown of the collinear approximation in the DGLV formula by considering the required ordering $k^+\gg k^-$.  From Eq. \ref{Eqn:momenta}, $k^+\simeq2xE$ and $k^-\simeq\textbf{k}^2/2xE$ we require
	\begin{equation}
	2xE  \gg \frac{\textbf{k}^2}{2xE} \quad 
	\Rightarrow \quad 2 \ngg \mathcal{O}(1),
	\end{equation}
	where the $\mathcal{O}(1)$ term ranges from $\sim\!\!1/2$ up to 2.  The lower limit of $1/2$ comes from considering the typical momentum fraction taken from the parent parton by the emitted gluon, $x_{typ}\sim\mu/E$ \cite{Horowitz:2009eb}; the upper limit of $2$ results from using $k_{max}=2xE$.  Thus the collinear approximation is violated for much of the phase space of the emitted gluon.
	
	Similarly, the large formation time approximation requires that
	\begin{equation}
	\mu  \gg \omega_i \sim \frac{1}{\tau} = \frac{\textbf{k}^2}{2xE} 
	\ngg \mu \times \mathcal{O}(1),
	\end{equation}
	where, again, the $\mathcal{O}(1)$ term ranges from $\sim \!\! 1/2$ up to 2.
	
	It is important to note that the large formation time assumption is a separate approximation from the collinear approximation; it is only when $\vert\textbf{k}\vert\sim\mu$ that the two approximations are equivalent.  	Nevertheless, and despite this \textit{a posteriori} understanding, the present calculation was performed making full use of the large formation time assumption.

	\section{\texorpdfstring{Sensitivity to small $ \Delta z $}{}}
	
	There is a lack of theoretical control over the physics of small $ \Delta z $ in heavy ion collisions, including, but not limited to, the factorization of a hard parton in the presence of early time strong fields and the thermalization of the medium.  It is therefore valuable to investigate the sensitivity of the energy loss to the details of small $ \Delta z $ physics.  In this section, we investigate the small $ \Delta z $ robustness of the energy and mass dependence of the correction term, seen in the previous section.

	\subsection{Distribution of scattering centers}
	
	The energy loss formula in Eq.~\ref{eq:ELossFull} contains an integral over the distribution of scattering centers $ \rho(\Delta z) $, which one is free to choose.  The original DGLV calculation assumed an exponential distribution, motivated by an attempt to mimic a rapidly expanding medium.  We have already alluded to the fact that an exponential distribution biases toward small separation distances, an effect which is exaggerated in small systems.  In order to counter this bias and to further explore the sensitivity of the energy loss calculations to early time dynamics, it is useful to consider other distributions of scattering centers.

	As a first step, and in order to avoid the complications of biasing toward small separation distances, we start our investigation by considering, as has been done by \cite{Djordjevic:2009cr}, a step function distribution of scattering centers.  This function is a properly normalized Heaviside-theta function which distributes the scattering centers evenly for all $ 0\leq \Delta z \leq L $, and we will refer to it as the ``full step function'' (abbreviated to ``F'' where necessary) for reasons that will become clear as we start to consider modifications of the simple step function.

	Secondly, one might attempt to investigate the sensitivity of the relative energy loss to small separation distances by imposing a lower cut-off for $ \Delta z $. The medium is modeled by Gyulassy-Wang potentials (see Eq.~\ref{eq:GWPotential}) that explicitly require small $ \sfracmu $, setting a convenient scale for what ``small $ \Delta z $'' might mean.  We, therefore, propose a modification of Eq.~\ref{eq:ELossFull} so that $ \rho(\Delta z) $ is a properly re-normalized truncated step function in which the scattering centers are evenly distributed between $ \sfracmu\leq \Delta z\leq L $.  The re-normalization needs to be such that the probability of scattering between $ \sfracmu $ and $ L $ is one.  Physically, in this instance, we envision producing a hard parton (its production having been properly factorized) before the medium has thermalized.  The parton might, therefore, travel a short $\Delta z \lesssim\sfracmu $ distance through an unthermalized medium that has not yet formed quasiparticle scattering centers, keeping in mind that, since we consider first order in opacity, we require exactly one scattering to take place.  We will call this distribution the ``truncated step function'' (abbreviated to ``T'' where necessary).
	
	Thirdly, recall that pQCD energy loss formalisms assume that the production of the hard parton may be factorized from its propagation through the medium.  The production mechanisms for hard partons in the presence of strong fields, and the scales on which they occur, have not yet been fully explored.  However, the present calculation is performed within the framework of DGLV energy loss, which is a static brick problem, and therefore does not take into account the details surrounding the production of hard partons.  
	In order to investigate this lack of information surrounding the factorization of the hard production processes, we propose a distribution of scattering centers which prohibits any energy loss from occurring close to the production.  We impose such a cut-off on the energy loss by applying a unit step function to the energy loss formula, while employing the full unit step distribution of scattering centers.  In practice, this truncation of the energy loss is implemented by truncating the unit step function distribution of the scattering centers without renormalizing, so that the probability of scattering is constant for $ 0\leq \Delta z\leq L $, but the energy loss is zero for $ \Delta z\leq \sfracmu $.
	
	Physically, this distribution is intended to mimic a hard parton that, having not yet formed properly, will not lose energy for some distance ($ 0\leq\Delta z\leq\sfracmu $) even if it should encounter a scattering center. We will call this distribution the ``truncated un-renormalized step function'' (abbreviated to ``TU'' where necessary).
	
	\begin{figure}
		\centering
		\includegraphics[scale=0.75,trim=1cm 0 0 0]{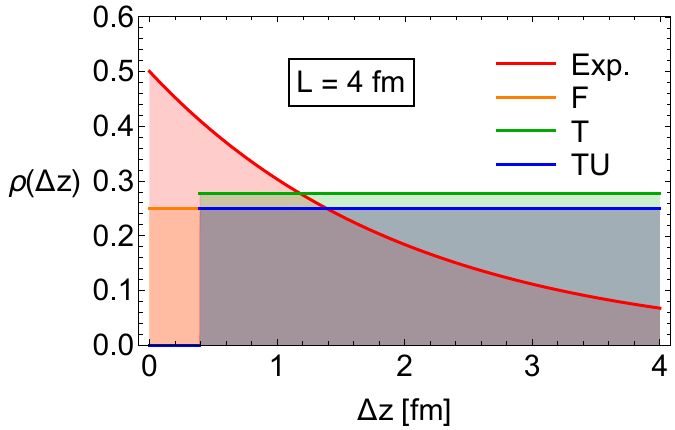}
		\caption{(Color online) (Grayscale coloring: `Exp' is the highest curve at $ \Delta z = 0 $, `F' is the lightest of the lower horizontal lines with $ \rho(\Delta z) \sim 0.25 $, `T' is the highest horizontal line with $ \rho(\Delta z) \sim 0.29 $, and `TU' the darkest of the lower horizontal lines with $ \rho(\Delta z)\sim 0.25 $.) The four different options for $ \rho(\Delta z) $, the distribution of scattering centers, discussed in the present article and described in \cref{eq:Exp,eq:F,eq:T,eq:TU}, as a function of $ \Delta z $.  In this particular set of curves we have chosen a system with $ L=4 $ fm. \label{fig:DistributionsPlot}}
	\end{figure}

	In summary, the four scattering center distribution functions we consider in this article are given by
	\begin{align}
	\rho_{\text{exp}}(\Delta z)	&=\frac{2}{L}\,\text{exp}\left(-\frac{2\Delta z}{L} \right)\label{eq:Exp}\\
	\rho_{\text{F}}(\Delta z)	&=\frac{1}{L}\Theta\left(L-\Delta z\right)\label{eq:F}\\
	\rho_{\text{T}}(\Delta z)	&=\frac{1}{L-\sfracmu}\Theta\left(\Delta z-\sfracmu \right)\Theta\left(L-\Delta z\right)\label{eq:T}\\
	\rho_{\text{TU}}(\Delta z)	&=\frac{1}{L}\Theta\left(\Delta z-\sfracmu\right)\Theta\left(L-\Delta z\right)\label{eq:TU},
	\end{align}
	and are shown in Fig.~\ref{fig:DistributionsPlot} for a brick of $ L=4 $ fm.
	
	\begin{figure}
		\centering
		\includegraphics[scale=0.65]{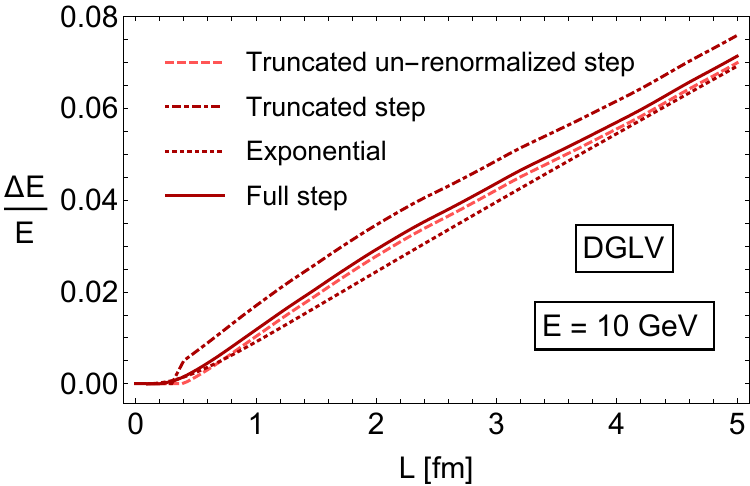}
		\caption{The relative DGLV energy loss of a bottom quark without small separation distance correction, as computed using the four different distribution functions for the scattering centers described in Eqs.~\ref{eq:Exp}~-~\ref{eq:TU}.  This plot is to be compared directly to Fig.2 in \cite{Djordjevic:2003zk}.  Note that the relative energy loss when using the truncated step function (dot-dashed curve) does not smoothly go to zero as $ L\rightarrow \sfracmu $ due to the normalization factor in Eq.~\ref{eq:T}. \label{fig:DGLVtoLBottomPlot}}
	\end{figure}
	
	In Fig.~\ref{fig:DGLVtoLBottomPlot} we show, having chosen the same parameters as were used in \cite{Djordjevic:2003zk}, the DGLV relative energy loss of a bottom quark without small separation distance correction, utilizing the four scattering center distributions described in Eqs.~\ref{eq:Exp}~-~\ref{eq:TU}.  Fig.~\ref{fig:DGLVtoLBottomPlot} is to be compared directly to Fig.~2 in \cite{Djordjevic:2003zk}.  In Fig.~\ref{fig:DGLVtoLBottomPlot} it can be seen, and we will show in Sec.~\ref{sec:EarlyTimeSensitivity} again, that the original DGLV term is not particularly sensitive to the choice of distribution.  The distribution with the biggest difference in energy loss is the truncated (renormalized) step.  This distribution biases the scatterings to larger $ \Delta z $, causing the bias toward larger energy loss.
	Note that the relative energy loss when using the truncated step function (dot-dashed curve in Fig.~\ref{fig:DGLVtoLBottomPlot}) does not smoothly go to zero as $ L\rightarrow 1/\mu $ due to the normalization (as $ L\rightarrow\sfracmu $, the normalization diverges like $ (L-\sfrac{1}{\mu})^{-1} $).  The almost complete lack of sensitivity to the differnces in the other distributions can be understood from formation time effects and a subtle cancellation of terms discussed further in Sec.~\ref{sec:EarlyTimeSensitivity}.

	\subsection{\texorpdfstring{Energy and mass dependence at small $ \Delta z $}{}}
	
	\begin{figure*}
		\centering
		\includegraphics[scale=0.55,trim=2cm 0cm 2cm 0cm]{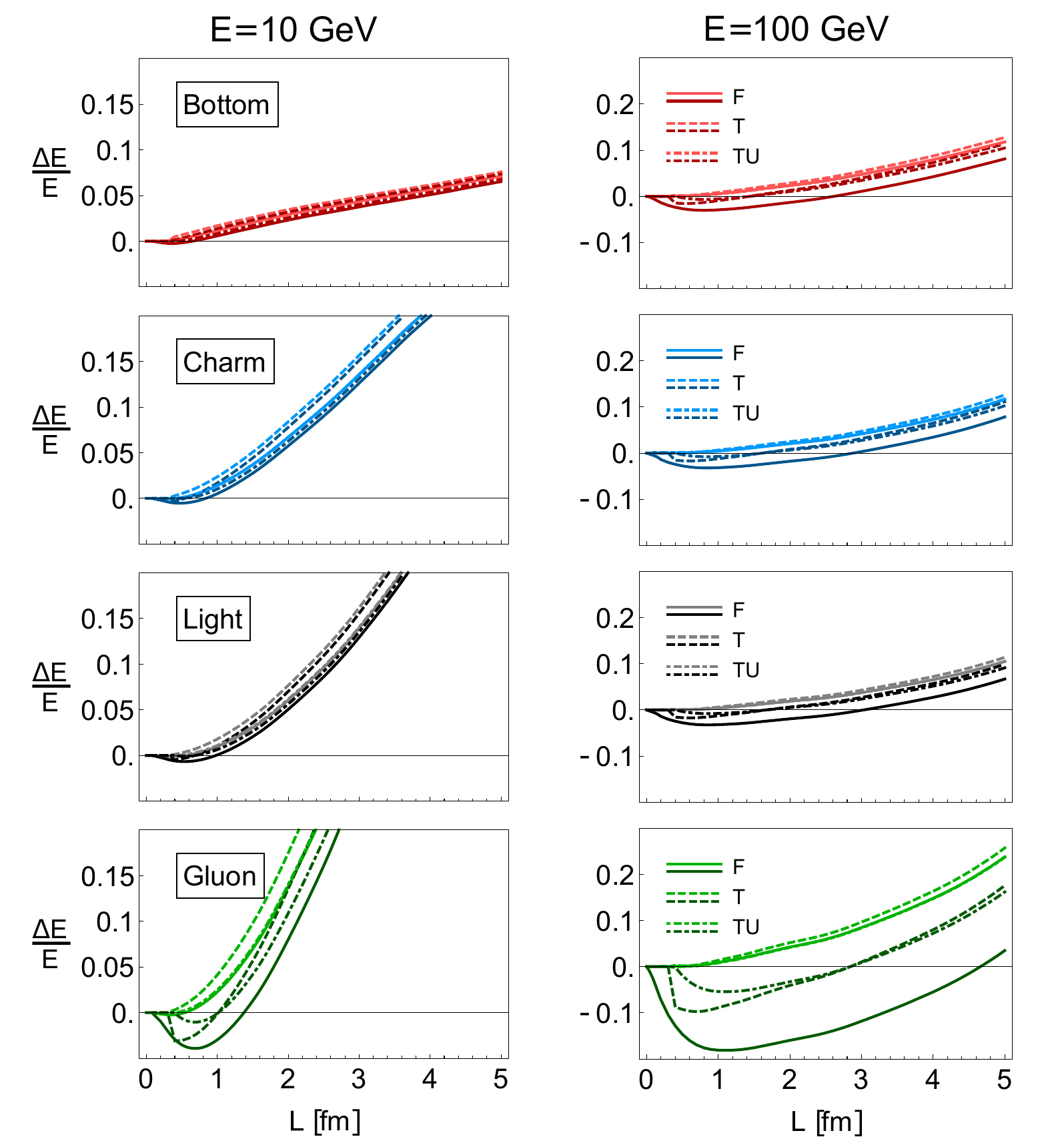}
		\caption{The relative energy loss of four different parent parton flavors (organized in rows), for parent parton energies of $ E=10 $ GeV (left column) and $ E=100 $ GeV (right column), for both the original large separation distance DGLV result (light curves) and the present all separation distance result (dark curves), as computed using the full step function (solid curves), the truncated step function (dashed curves) and the truncated un-renormalized step function (dot-dashed curves).\label{fig:GridOf8PaperPlot}}
	\end{figure*}

	The sensitivity of Eq.~\ref{eq:ELossFull} to the choice of $ \rho(\Delta z) $ may be further investigated by considering more closely the sensitivity of the flavor and energy dependence of the correction to the choice of $ \rho(\Delta z) $.
	To this end, we present a number of plots in Fig.~\ref{fig:GridOf8PaperPlot}, showing the relative energy loss $ \Delta E/E $ for four different parent parton flavors (grouped in rows) at $ E=10 $ GeV (left column) and $ E=100 $ GeV (right column).  All of these plots show the curves produced by using the full step function (solid curves), the truncated step function (dashed curves), and the truncated un-renormalized step function (dot-dashed curves)\footnote{We have not included the exponential distribution here as it lends little to the present discussion.}, for both the large separation distance DGLV result (light curves) and the present all separation distance result (dark curves).

	By considering the dark curves in Fig.~\ref{fig:GridOf8PaperPlot}, showing the all separation distance result Eq.~\ref{eq:ELossFull}, it is clear that the correction term is sensitive to the choice of distribution of scattering centers.  We investigate the reasons for this sensitivity in section \ref{sec:EarlyTimeSensitivity}.  The dominance of the correction term at high energies (right-hand column) is described in section \ref{sec:EnergyDependence}.  One may understand the crossover of the truncated step function (T) and truncated un-renormalized step function (TU) curves (most easily seen in the $ E=100 $ GeV plots, but also present in the $ E=10 $ GeV plots) as a result of the fact that the T distribution biases toward larger separation distances so that, at larger $ L $, the characteristic $ L^2 $ dependence of the DGLV energy loss overpowers the $ L^0 $ dependence of the correction term  at a smaller $ L $ (see Eqs.~\ref{eq:AsympNLO} and \ref{eq:AsympLO}).
	
	The column on the right in Fig.~\ref{fig:GridOf8PaperPlot} also clearly shows that, at $ E=100 $ GeV, the mass dependence of the relative energy loss of the quarks disappears.  This may be understood by recalling that the momentum of the radiated gluon \textbf{k} is integrated over from 0 to $ 2x\,(1-x\, E) $, so that masses in both the momentum prefactors and the formation times in Eq.~\ref{eq:ELossFull} are overpowered by the $ \textbf{k}^2 $ at large $ E $.

	\begin{figure*}
		\centering
		\includegraphics[scale=0.9]{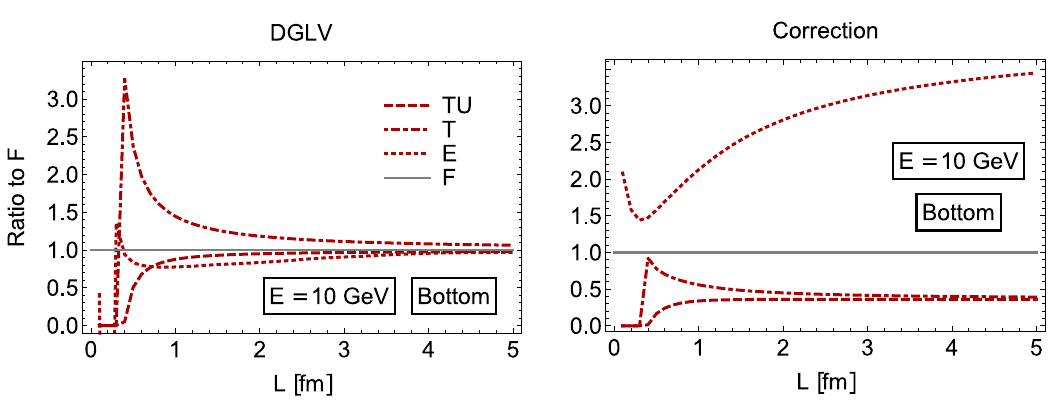}
		\caption{The ratio of the relative energy loss as computed using various different scattering center distributions (truncated un-renormalized step function in dashed curves, truncated step function in dot-dashed curves and exponential in dotted curves) to that computed using the full step function distribution, for an $ E=10 $ GeV bottom quark as a function of the size $ L $ of the brick, for the DGLV term (left-hand panel) and the small separation distance correction term (right-hand panel).  This ratio is unity for an energy loss formula that is insensitive to the physics of small separation distances.\label{fig:DistributionRatios}}
	\end{figure*}	
	
	In order to further quantify the sensitivity of the energy loss to early time physics, we plot in Fig.~\ref{fig:DistributionRatios} ratios of relative energy loss computed using three different scattering center distributions (truncated un-renormalized step function in dashed curves, truncated step function in dot-dashed curves and exponential in dotted curves) to the relative energy loss computed using the full step function, for the large separation distance DGLV result on the left, and the present small separation distance correction on the right, all for an $ E=10 $ GeV bottom quark.  This ratio is unity for an energy loss formula that is insensitive to the physics of $ \Delta z\lesssim\sfracmu $.  One immediately notes that, while the DGLV results all tend toward one, the correction term's sensitivity to the early time dynamics is persistent even at large $ L $.  One can see that, compared to the unit step, varying the scattering center distribution leads to up to a factor of two reduction, or factor of four enhancement, of the correction term.  Although not presented here, we have computed the ratio of E to F for $ L $ up to 100 fm and find that this ratio asymptotically approaches $\sim\!4 $, although the validity of the first order in opacity expansion for such large path lengths becomes questionable.  The large deviation of T away from F for the DGLV result at small $ L $ is due to the normalization of T, as well as a very small energy gain and subsequent division by zero for small path lengths. 
	
	\begin{figure*}
		\centering
		\includegraphics[scale=0.5,trim=1.2cm 0cm 1cm 0cm]{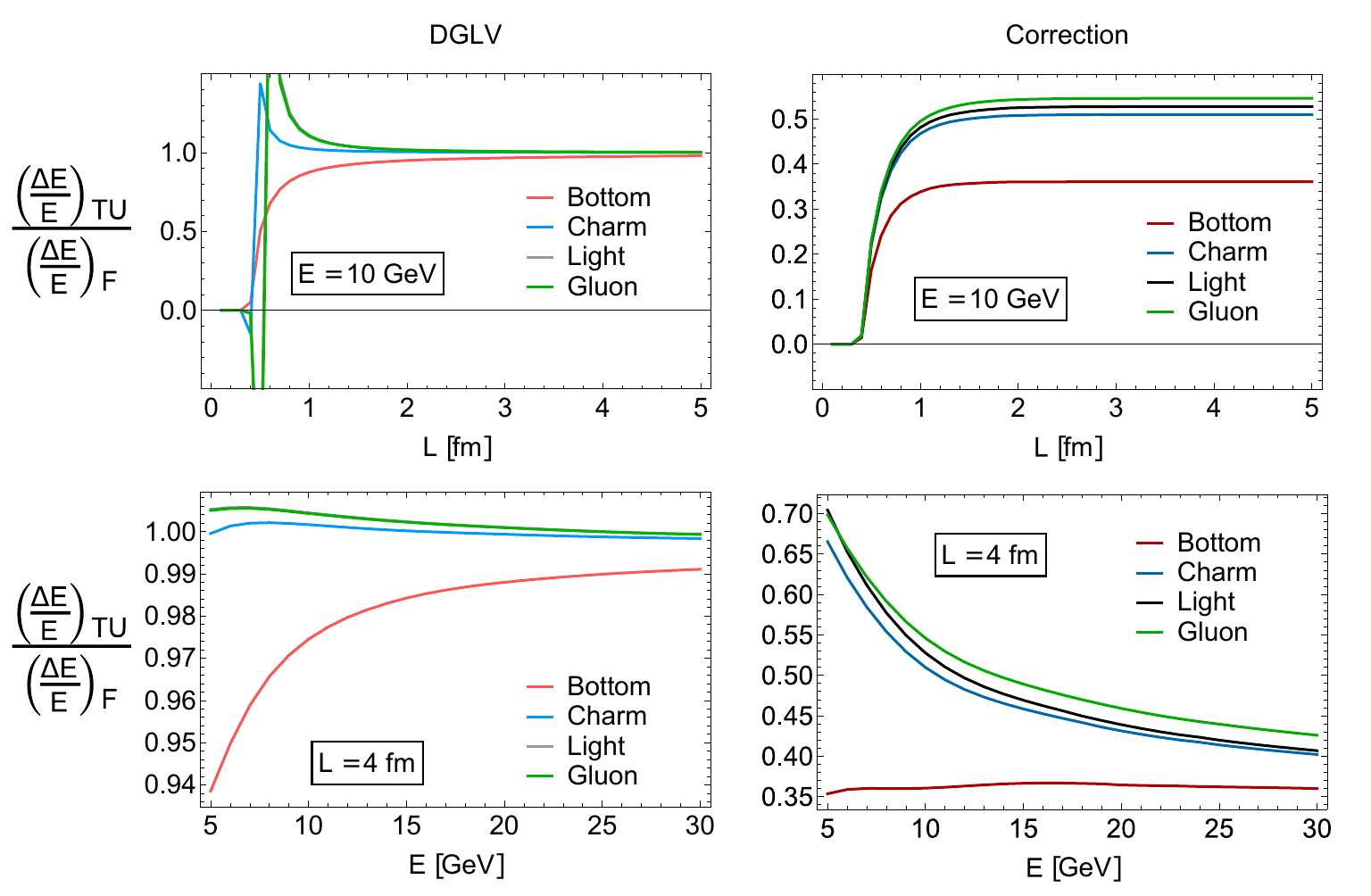}
		\caption{(Color online) (Grayscale coloring guide: In all four plots, except below $ L\sim 1 $ in the top left plot, the curves, from bottom to top, represent `Bottom', `Charm', `Light', `Gluon', respectively, with the `Light' and `Gluon' curves overlapping in the left column.) The ratio of the relative energy loss as computed using the truncated un-renormalized step function to that computed using the full step function.  This ratio is shown in the top row as a function of the length of the brick for parent partons with $ E=10 $ GeV and in the bottom row as a function of the energy of the parent parton moving through a brick of $ L=4 $fm, for DGLV (left column) and for the correction (right column).  This ratio is unity for an energy loss distribution that is insensitive to the physics of $ \Delta z\lesssim\sfracmu. $  \label{fig:DGLVHKtoLFTUPapPlot}}
	\end{figure*}

	We may investigate the mass and energy dependence of the differences between scattering center distributions even further by considering the plots presented in Fig.~\ref{fig:DGLVHKtoLFTUPapPlot}, where we plot the ratio of the relative energy loss as computed using the truncated un-renormalized scattering center distribution to that computed using the full step function, for the DGLV result (left column) and the correction term (right column).  The plots in the left column of Fig.~\ref{fig:DGLVHKtoLFTUPapPlot} show that the insensitivity of the DGLV result to small system dynamics is independent of both mass and energy, particularly for $ L\gtrsim 1 $fm.  On the other hand, differences of a factor of 2 persist to all path lengths for the correction term. 
	The length dependent DGLV ratio in the top left hand corner of Fig.~\ref{fig:DGLVHKtoLFTUPapPlot} exhibits some fluctuant behavior at small $ L $ for some flavors, due, as in Fig.~\ref{fig:DistributionRatios}, to numerical division by zero. 
	
	The correction term's sensitivity to small $ \Delta z $ physics is also seen to be mass dependent in Fig.~\ref{fig:DGLVHKtoLFTUPapPlot}, with the bottom quark most affected by the truncation of the scattering center distribution.  Although the overall mass dependence  of the relative energy loss at low energies is mostly due to the mass dependence of the propagators in Eq.~\ref{eq:ELossFull} (discussed in Sec.~\ref{sec:MassOrdering}), the ratio of relative energy losses divides out any mass dependence that is not coupled to the separation distance.  We may therefore understand the mass dependence of the ratio shown in Fig.~\ref{fig:DGLVHKtoLFTUPapPlot} from a formation time perspective:  Consider the formation time of a gluon radiated off a parent parton with mass $ M $, given by 
	\begin{equation}
	\tau_{f}	\equiv\frac{2 x E}{\textbf{k}^2+x^2M^2}.
	\end{equation}
	The high mass of the bottom quark will then give the bottom quark the shortest radiated gluon formation time.  The shorter the formation time, the more sensitive will the parton be to early time physics.
	One expects such a mass dependence to disappear at high energies, and indeed, the sensitivity of the relative energy loss to the choice of distribution appears to converge for the different quark masses at high energies, as seen in the bottom left plot of Fig.~\ref{fig:DGLVHKtoLFTUPapPlot}.  Naively one might expect all the quarks to appear massless (and so to see the ratio in Fig.~\ref{fig:DGLVHKtoLFTUPapPlot} converge to the light quark result rather than that of the bottom quark at high energies).  For the DGLV result (bottom left in Fig.~\ref{fig:DGLVHKtoLFTUPapPlot}), this intuition holds because the DGLV result is insensitive to small separation distance physics.  
	On the other hand, because of the correction term's sensitivity to small separation distance dynamics, and since higher energies result in shorter radiation formation times, the curves in the bottom right panel of Fig.~\ref{fig:DGLVHKtoLFTUPapPlot} tend toward the bottom quark result, since it is the bottom quark that already has the shortest formation time.

	\subsection{\texorpdfstring{Origins of small $ \Delta z $ sensitivity}{}}\label{sec:EarlyTimeSensitivity}
	
	\begin{figure*}
		\centering
		\includegraphics[scale=0.45]{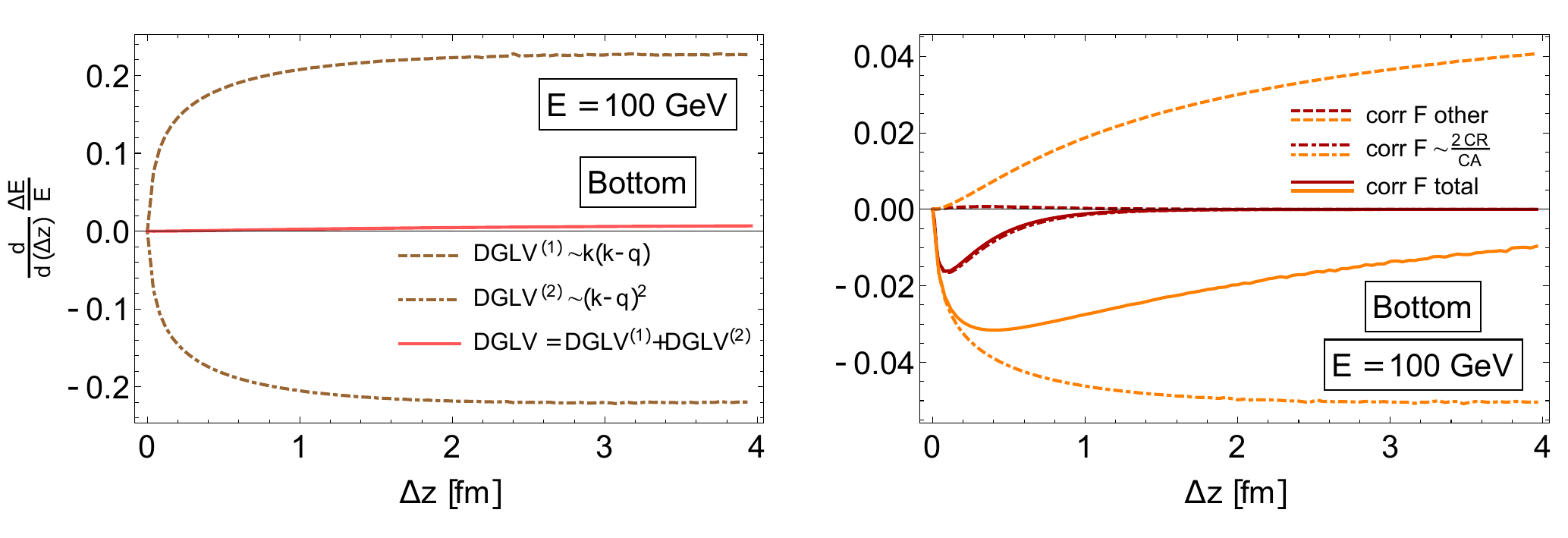}
		\caption{(Color online) The DGLV term (left-hand panel) and correction term (right-hand panel) contributions to the $ d(\Delta z) $ differential of the relative energy loss of an $ E = 100 $ GeV bottom quark moving through a brick of $ L = 4 $ fm, showing the contributions from individual terms in Eq.~\ref{eq:ELossFull}. in the panel on the left, the dashed and dot-dashed curves show the two terms in brackets in the second line of Eq.~\ref{eq:ELossFull}, while the solid curve shows their sum (the full DGLV result). In the right-hand panel the dashed curves show the sum of the two terms that cancel in the high energy limit (see Sec.~\ref{sec:EnergyDependence} for details), the dot-dashed curves show the color carrying term and the solid line their sum.  For the correction term in the right-hand panel, the red (darker) curves show the full result while the orange (lighter) curves show what the contributions to the correction term would be without the $ \text{exp}(-\mu\Delta z) $ factor (see Sec.~\ref{sec:EarlyTimeSensitivity} for details).\label{fig:SplitPlot}}
	\end{figure*}
	In Eq.~\ref{eq:ELossFull} we see that both the DGLV terms and the correction terms contain formation times; i.e.,\ the terms are proportional to cosines of argument $ \omega_i \Delta z $ such that $ \Delta E_{DGLV} $ and $ \Delta E_{corr} $ go to zero for $ \Delta z\lesssim1/ \omega_i $.  It is therefore difficult to understand the sensitivity of the correction term to early time physics, in conjunction with the \textit{insensitivity} of the DGLV term, from a formation time perspective.	Investigating the DGLV term further numerically, one finds a subtle cancellation that occurs in the DGLV term that does not occur in the correction term.  In the DGLV term, the two terms in the brackets in the second line of Eq.~\ref{eq:ELossFull} ($ DGLV^{(1)}\sim k(k-q) $ and $ DGLV^{(2)}\sim (k-q)^2 $, so that $ DGLV =DGLV^{(1)}+DGLV^{(2)}$) are very large but almost equal in magnitude and opposite in sign.  As such, the two contributions to the DGLV term cancel almost identically, which may be seen in the left hand panel of Fig.~\ref{fig:SplitPlot}, where we plot the contributions from $ DGLV^{(1)}$ and $ DGLV^{(2)} $ separately, along with their sum, for an $  E=100 $ GeV bottom quark.  
5	
	No such cancellation occurs in the correction term, a fact we already alluded to in Sec.~\ref{sec:EnergyDependence} where we found that two of the three terms in the correction cancel, while the color triviality breaking term remains and is responsible for the bulk of the contribution.  To illustrate the dominance of the color triviality breaking term in addition to the cancellation of the remaining terms of the correction, we present the red curves in the right-hand panel of Fig.~\ref{fig:SplitPlot}, showing the contributions from the two terms that cancel in the high energy limit, the color triviality breaking term, and their sum, for an $ E=100 $ GeV bottom quark.  One can see in the red curves of the right-hand panel of Fig.~\ref{fig:SplitPlot} that the color triviality breaking term controls the correction term's energy loss.  Therefore, the DGLV term's insensitivity to the small separation distance physics is due to \textit{both} the destructive LPM effect \textit{and} this subtle cancellation effect, while the absence of such a cancellation in the correction term contributes to the 
	correction term's sensitivity to small $ \Delta z $.
\begin{flushright}
	
\end{flushright}	
	Additionally, the correction contains a factor of $ \text{exp}\left(-\mu\Delta z\right) $, which plays the part of suppressing contributions to the correction term from $ \Delta z\gtrsim \sfracmu $, enforcing a strong dependence on the physics of $ \Delta z\lesssim\sfracmu $. In order to understand the role of the exponential factor in the sensitivity of the correction term to the small separation distance physics, we present the orange curves in the right-hand panel of Fig.~\ref{fig:SplitPlot}, which show the same three terms of the correction term as are shown in the red curves, but without the factor of $ \text{exp}\left(-\mu\Delta z\right) $. It is clear that, upon integrating over $ \Delta z $, the bulk of the contributions to the integral comes from the region $ \Delta z \lesssim\sfracmu $ due to the presence of the $ \text{exp}\left(-\mu\Delta z\right) $ factor.
	
	\section{Conclusions}\label{sec:Conclusions}
	
	The original DGLV derivation of the energy loss of a hard, potentially massive parton via radiation (of potentially massive quanta), while traversing a static brick of weakly coupled QGP, assumed a large path length for the parent parton.  In this article, we generalized the first order in opacity of DGLV by including the short path length terms that were neglected in the original derivation. 
	We have thus analytically derived a small separation distance correction to the first order in opacity of DGLV.  Our result constitutes an important step toward the understanding of partonic energy loss in small colliding systems. 
	
	The main result of our article is the all separation distance first order in opacity energy loss formula Eq.~\ref{eq:ELossFull}.  In our derivation we retained the scale ordering of $ \sfracmu\ll\lambda_{mfp} $, justifying the use of the Gyulassy-Wang model, as well as the soft and collinear assumptions, and we have retained the usual assumption of large formation time. 
	We found that the majority of terms that are exponentially suppressed under the large separation distance assumption are additionally suppressed under the large formation time assumption at the amplitude level, meaning that only two diagrams out of twelve contribute to the small separation distance correction.
	We performed an extensive numerical analysis of the correction term and found that, surprisingly, the correction term dominates over the original DGLV result at high energies.  This energy dependence may be understood from an asymptotic analysis that revealed an $ E\log E $ energy dependence of the correction term, in contrast to the $ \log E $ dependence of the large separation distance DGLV.  We further found that the correction term depends on the distance traveled through the medium as $ L $ for small $ L $ and $ L^0 $ for large $ L $ (that is, the correction is independent of $ L $ for large $ L $), again deviating from the $ L^2 $ dependence of the DGLV term.  Therefore, the effects of the correction term persist to arbitrarily large paths.  Interestingly, the correction term also breaks color triviality.

	Naively one might expect aspects such as the factorization of the production of the hard parton from the scattering, the behavior of a Debye screened scattering center near the edge of a thermalized medium, etc., to play a role in small system modeling.  In order to explore the effect of the physics of small systems and early times on our correction term, we proposed a number of distributions of scattering centers, attempting to take into account the factorization of the production of the hard parton from its propagation through a medium, as well as the formation of that medium.  We showed that the short separation distance correction is sensitive to early time physics explored by these distributions, while the original large separation distance DGLV result is not.  
	We found that the DGLV term's insensitivity to the physics of small $ \Delta z $ is due to both the known formation time physics and a subtle cancellation of terms.  This cancellation does not persist in the correction term, which accounts for the sensitivity of the correction term to small $ \Delta z $ physics.
	
	Our derivation revealed that the formation time of a gluon radiated off a hard parent parton is of crucial importance.  Already at the amplitude level of the all separation distance derivation, we found that the naive application of the large formation time assumption leads to a dramatic reduction of terms present in the correction.  Phenomenologically, one might ask if it is justifiable to include large formation times while considering short path lengths.  It's important to remember that the path length is the distance within which a scattering of the parent parton or radiated gluon occurs; there is no requirement that the radiated gluon forms within the ``brick'' of medium we consider.  We also demonstrated that the large formation time assumption is violated for much of the phase space of the emitted radiation.   Since all energy loss formalisms, DGLV, BDMPS-Z-ASW, AMY, and HT (see \cite{Majumder:2010qh} and references therein) exploit the large formation time approximation, we are faced with a need to assess the applicability of the large formation time assumption in any description of energy loss.  While the influence of the assumption of collinearity was relatively easy to quantify across formalisms by simply varying the maximum allowable perpendicular momentum of the emitted gluon, estimating the importance of the large formation time approximation will likely be a challenge.  Similarly, deriving expressions that do not rely on either the collinear or large formation time approximations is formidable.  
	
	The physics of formation times is also relevant to the mass ordering of the energy loss at high parent parton energies.  However, we found that the mass ordering is additionally subject to competing effects from the massive propagator, so that the mass dependence of the relative energy loss at low energies is dominated by the propagator.
	
	Our results show that, if one is to consider a system in which the separation distances are on the order of the Debye screening length, one will have to understand the in-medium production mechanisms as well as the nature of a Debye screened scattering center near the edge of a thermalized medium, in addition to the validity of the large formation time assumption in small systems.  Due to these large uncertainties, the quantitative effect of the correction on observables is unclear.  Further, the lack of theoretical control over these assumptions calls into doubt the quantitative extraction of medium parameters through the use of jet quenching \cite{Burke:2013yra}.  We leave addressing these issues for future work.
	
	\section{Acknowledgments}
	
	The financial assistance of the National Research Foundation (NRF) and the German Academic Exchange Service
	(DAAD) toward this research is hereby acknowledged. Opinions expressed and conclusions arrived at are those of the
	authors and are not necessarily to be attributed to the NRF
	or DAAD. The authors thank the SA-CERN Collaboration,
	the University of Cape Town, and the National Institute for
	Theoretical Physics (NITheP) for their generous support. The
	authors wish to thank both CERN and BNL for the generous
	hospitality extended to them during the completion of part
	of this work. I.K. additionally acknowledges partial support
	from the U.S. Department of Energy, Office of Science, Office
	of Nuclear Physics, under Contract No. DE- SC0012704,
	as well as support from iThembaLABS. Finally, the authors
	thank Magdalena Djordjevic, Miklos Gyulassy, Ulrich Heinz,
	Sangyong Jeon, Urs Wiedemann, and Carlos Salgado for
	valuable discussions.

\bibliography{InspireBib}

\bibliographystyle{apsrev4-1}

\end{document}